\documentstyle[aps,multicol]{revtex} 

\def\tup{\begin{array}{c}\uparrow\\[-1mm] -\end{array}}
\def\tdown{\begin{array}{c}\downarrow\\[-1mm]-\end{array}}
\def\bup{\begin{array}{c}-\\[-1mm]\uparrow\end{array}}
\def\bdown{\begin{array}{c}-\\[-1mm]\downarrow\end{array}}

\def\bpi{\pi\!\!\!\! \pi}

\def\braket#1{\left|#1\right>}

\begin{document}
\draft
\title
  {Theory of electrons with orbital degeneracy}
\author
  {You-Quan Li${}^{1,2}$ and Ulrich Eckern${}^1$}
\address
  {${}^{1}$ Institut f\"ur Physik, Universit\"at Augsburg, 
   D-86135 Augsburg, Germany\\
   ${}^{2}$ Zhejiang Institute of Modern Physics, Zhejiang University, 
   Hangzhou 310027, China}

\date{Received: Nov. 16, 1999; revised: July 7, 2000}

\maketitle

\begin{abstract}
The Hubbard model for electrons with orbital degeneracy is shown
to have an underlying $SU_d(4)$ symmetry of spin-orbital double. 
A hidden charge $SU_c(4)$ symmetry is exposed and 
an extended Lieb-Mattis transformation which maps these two symmetries 
into each other is given.  On the basis of elementary degenerate
perturbative theory, it is shown that the system with strong repulsive
coupling is equivalent to a SO(6) Heisenberg magnet at half-filling
and a SU(4) one at quarter-filling.
The band is half-filled at all temperature for $\mu=3U/2$.
The features of ground state 
and low-lying excitations in one dimension are indicated according
to exact solutions. Various possibilities of symmetry breaking
are also given.
\end{abstract}



\begin{multicols}{2}

\section{Introduction}\label{sec:introduction}

There has been much interests in the studies on correlated electrons 
in the presence of orbital degree of freedom\cite{Nagaosa} 
because the orbital degree of freedom plays an important role in
understanding the phenomena, such as metal-insulator transitions,
high-temperature superconductivity and colossal magneto-resistance. 
The orbital degree of freedom is relevant to many transitional 
metal oxides \cite{McWh,Word,Bao,Pen,Feiner}.
It may be also relevant to some $C_{60}$ materials
\cite{Arovas} and samples of artificial quantum dot arrays \cite{Marston}.
For a theoretical understanding of the observed unusual properties, 
a SU(4) theory describing spin systems with orbital degeneracy was
proposed \cite{LiMSZ}. There were also numerical \cite{Ueda} 
and perturbative \cite{Azaria} studies of 1-dimensional models for
these systems. The ground-state phase diagrams for the system with
a symmetry breaking of $SU(4)\rightarrow SU(2)\times SU(2)$ were discussed
\cite{Azaria,IQAffleck}. Experimentally, the phase separation
\cite{phasesep} was observed. Due to the rapid 
developments in experiments where the metal ions has orbital 
degeneracy in addition to spin degeneracy,
a theoretical study of such system by taking account of the kinetic terms 
caused by nearest neighbor hopping becomes indispensable.

In this paper we study a Hubbard-type model for electrons with orbital
degeneracy. In Sec. \ref{sec:underlying}, we show that the model 
has an underlying SU(4) symmetry of spin-orbital double. 
The spin and orbital operators are related to the SU(4) generators,
which will be helpful for further studies on the magnetization.
A hidden charge SU(4) structure is exposed in Sec. \ref{sec:hidden}.
An extended Lieb-Mattis transformation which maps those two SU(4) 
symmetries into each other is also presented. 
From a basic relation derived from particle-hole transformation, 
we show that the band is half-filled at all temperature when the chemical
potential equals $3U/2$.
In Sec. \ref{sec:partial},
two kinds of ``partially negative $U$'' models are introduced and analyzed
according to the strategy of \cite{Emery}. 
Three basic excitation modes are show to exist in the spin-orbital sector.
In Sec. \ref{sec:effect}, with the help of the partially attractive models,
we study the repulsive large $U$ model on the basis of elementary degenerate
perturbative theory.
It is shown that the effective Hamiltonian with strong
repulsive coupling at half-filling is equivalent to the Hamiltonian of 
SO(6) Heisenberg model, and that at quarter-filling is equivalent to 
the one of SU(4) Heisenberg model.
In Sec. \ref{sec:breaking}, several possibilities of SU(4) symmetry breaking
are given.
In Sec. \ref{sec:discussions}, we summerize the main results and discuss
the agreement with exact solutions 
in one dimension by describing the features of  ground state and 
low-lying excitations.

\section{Underlying SU(4) symmetry}\label{sec:underlying}

We consider electrons with doubly orbital degeneracy.
The spin components are denoted by up $(\uparrow)$ and down $(\downarrow)$,
the orbital components by top and bottom. The four possible states of
electrons are
\begin{eqnarray} 
|1>=|\tup >, \;\;\;&  
|2>=|\tdown >, \nonumber\\[0mm] 
|3>=|\bup >, \;\;\;& 
|4>=|\bdown >.
\end{eqnarray}
We use $1, 2, 3,$ and $4$ to label these states from now on. 
Let us consider the Hamiltonian of  electrons with two-fold
orbital-degeneracy  on a lattice
\begin{equation}
H=-t\sum_{\stackrel{a}{<x,x'>}}C_a^+(x) C_a(x')
          +\sum_{\stackrel{a < a'}{x}}U_{a a'}n_a(x)n_{a'}(x).
\label{eq:Hamiltonian}
\end{equation}
where $x$'s identify the lattice site, $a, a'= 1, 2, 3, 4$ 
specify the spin and orbital as defined in the above.
The $C_a^+(x)$ creates a fermion of state $\braket{a}$ located
at $x$ site and $n_a(x)$ is the corresponding number operator. 
Eq. (\ref{eq:Hamiltonian}) is the Hamiltonian for four-component systems,
and there were various discussions on multi-component Hubbard
model in one dimension\cite{Schlottmann,Choy}. 
We remark that the four-component Hamiltonian
can also describe either a spin-$3/2$ system \cite{spin3/2}
or a toy model of proton and neutron
system with on-site strong interaction. In the terminology of group theory,
the former forms a high-dimensional (here it is 4 dimensional)
representation of $A_1$ Lie algebra, while the later forms
the fundamental representation of $D_2$ Lie algebra. Whence, the physics that
Eq. (\ref{eq:Hamiltonian}) describes will be precise only when the 
representation space for the internal degree of freedom is specified. It 
refers to the spin and orbital in our present discussion.

We can verify that the Hamiltonian (\ref{eq:Hamiltonian}) 
with $U_{aa'}=U$ commutes with following 15 operators,
\begin{eqnarray} 
\,&O_m =\displaystyle\frac{1}{2}\sum_x 
   [C_m^+(x)C_m(x)-C_{m+1}^+(x)C_{m+1}(x)],\nonumber\\
\,&E_{\alpha_m}=\displaystyle\sum_x C_m^+(x) C_{m+1}(x),
    \hspace{2mm}\nonumber\\
\,&E_{-\alpha_m}=(E_{\alpha_m})^\dagger, \hspace{6mm} m=1, 2, 3,
\label{eq:su4spob}
\end{eqnarray} 
and additionally, 
$E_{\alpha_1+\alpha_2}=[E_{\alpha_1},E_{\alpha_2}]$,
$E_{\alpha_2+\alpha_3}=[E_{\alpha_2},E_{\alpha_3}]$,
$E_{\alpha_1+\alpha_2+\alpha_3}=[E_{\alpha_1+\alpha_2},E_{\alpha_3}]$,
$E_{-\alpha_1-\alpha_2}=(E_{\alpha_1+\alpha_2})^\dagger$,
$E_{-\alpha_2-\alpha_3}=(E_{\alpha_2+\alpha_3})^\dagger$,
and
$E_{-\alpha_1-\alpha_2-\alpha_3}
=(E_{\alpha_1+\alpha_2+\alpha_3})^\dagger$,
here $\alpha_m$'s stand for the simple roots of $A_3$ Lie algebra.
These operators precisely obey the commutation relations of $A_3$
Lie algebra \cite{Gilmore}. Whence the Hamiltonian  
(\ref{eq:Hamiltonian}) with $U_{aa'}=U$ is invariant 
under any global SU(4) rotation.
We denote this underlying symmetry of spin-orbital 
double by $SU_d(4)$.
There are several equivalent ways to write out the generators 
of the Lie algebra. 
We adopt the Chevalley basis because physical
quantities can be conveniently evaluated in this basis.
Because the $A_3$ Lie algebra is of rank three, 
i.e., three generators in its Cartan sub-algebra,  
there are three conserved quantum numbers that label the eigenstates. 
In terms of these $O_m$'s that commute to each other, the $z$-components
of both total spin $S^z_{tot}$ and total orbital $T^z_{tot}$ are give by
\begin{eqnarray} 
S^z_{tot} &=& O_1 + O_3, \nonumber\\
T^z_{tot} &=& O_1 +2O_2 +O_3.  
\label{eq:zcomponent}
\end{eqnarray}  
They are useful for evaluating magnetizations once
the ground state in the presence of magnetic field is solved.

\section{Hidden charge SU(4) symmetry}\label{sec:hidden}

In addition to the above symmetry, one may easily think of the $U(1)$
charge invariance\cite{Assaraf}. 
Moreover, there exists a larger hidden symmetry in present model 
on a bipartite lattice, we call charge $SU_c(4)$ symmetry. 
Their Chevalley bases are given by
\begin{eqnarray} 
\,&Q_m =\displaystyle\frac{1}{2}\sum_x
     [C_m^+(x)C_m(x)+C_{m+1}^+(x)C_{m+1}(x)-1],
        \nonumber\\
\,&F_{\alpha_1}=\displaystyle{\sum_x}e^{i \bf \bpi\cdot x} C_1^+(x) C_2^+(x),
       \hspace{7.6mm}\nonumber\\
\,&F_{\alpha_2}=\displaystyle{\sum_x}e^{i\bf \bpi\cdot x} C_2(x) C_3(x),
       \hspace{9.6mm}\nonumber\\
\,&F_{\alpha_3}=\displaystyle{\sum_x}e^{i\bf \bpi\cdot x} C_3^+(x) C_4^+(x),
     \;\; m=1, 2, 3, 
\label{eq:su4charge}
\end{eqnarray}
where ${\bf \bpi}=(\pi, \pi,...)$,
$Q_m$'s are generators of Cartan subalgebra of the $A_3$ Lie
algebra. The other generators are given by standard relations that
we demonstrated previously when observing the underlying $SU_d(4)$ symmetry. 
The charge $SU_c(4)$ symmetry is not only valid for the Hamiltonian
(\ref{eq:Hamiltonian}) at high energy scale $t\gg U$ but also
valid for a kind of on-site coupling which we are going to show. 
Considering
$U_{13}=U_{24}=-U$ while $U_{a b}=U$ for the other subscripts
and taking account of the chemical potential term in the Hamiltonian, 
we obtain 
\begin{eqnarray}
[H', F_{\alpha_m} ]&=&(-1)^m (2\mu -U)F_{\alpha_m},\nonumber\\
H'&=&H-\mu\displaystyle{\sum_{x,a}}n_a(x).
\label{eq:chempotential}
\end{eqnarray}
The commutators between $H'$ and $Q_m$'s always vanish. 
Apparently, the model has a charge $SU_c(4)$ symmetry when
$\mu=U/2$. The mentioned requirement for the sign of the on-site coupling
constants is unnecessary for the traditional Hubbard model which
has a hidden charge $SU(2)$ symmetry \cite{chargeSU2} because 
there is only one constant for coupling of spin up and spin down.

Eq. (\ref{eq:chempotential}) implies that the raising operators 
$F_{\alpha_m}$
of the charge $SU_c(4)$ create some pair of electrons to a given state. 
Precisely, $F_{\alpha_1}$ or $F_{\alpha_3}$ creates a double occupancy
of spin singlet, however,  $F_{-\alpha_2}$ creates a double occupancy
of spin triplet. These operators provide mappings between states in distinct
sectors of different electron numbers.  

There exists an extended Lieb-Mattis transformation:
\begin{eqnarray}
C_i(x)&\mapsto&e^{i{\bf \bpi\cdot x}}C_i^+(x),\;\;i=2,4,\nonumber\\
C_j(x)&\mapsto&C_j(x),\;\;j=1,3,
\end{eqnarray}
which maps $SU_c(4)$ into $SU_d(4)$ and vice versa.
The application of particle-hole transformation
gives rise to a basic relation \cite{Choy} for bipartite lattice:
\begin{equation}
E(N_a, U)=E(L-N_a, U)+3(N-2L)U,
\label{eq:basic}
\end{equation} 
where $L$ is the total number of lattice sites and $N$ 
the total number of electrons. Using this relation we can derive
a relation for the thermal average:
\begin{equation}
<\hat{N}>_{\mu,T}=4L-<\hat{N}>_{3U-\mu,T},
\end{equation}
where $\hat{N}=\sum_{x,a}n_a(x)$.
As a result, the band is half-filled at all temperature when $\mu=3U/2$.

\section{Partially attractive models}\label{sec:partial}

The magnitude and sign of the on-site coupling may vary 
from system to system.  For isotropic pure attractive coupling
$U_{ab}=U<0$, the unperturbed ground state has $N/4$ of the sites 
occupied by ``quaternarys''. The ground state is degenerate when 
$N/4<L$ because the energy $3NU/2$ is independent of which sites
are occupied. 
As a generalization of Cooper pair,
the quaternary might have abundant physical meanings.
It can form a SU(4) singlet for $U_{ab}=U$.  It can 
also form either two pairs being spin singlet but orbital triplet
or that being spin triplet but orbital singlet, 
or form a ``resonance'' state being alternations of them
depending on the symmetries remained in a general $U_{ab}$. 
For example, a triplet-pairing supper conductivity model was discussed 
\cite{Aligia} by a particular choice of $U_{ab}$.

We have particular interests in two kinds of 
partially attractive on-site couplings:
(i) $U_{13}=U_{24}=-U$ while $U_{ab}=U$ for the others;
(ii) $U_{12}=U_{23}=U_{13}=-U$ while $U_{14}=U_{24}=U_{34}=U$. 
Here $U<0$ for both cases.
Let us consider these two cases respectively. 

In case (i), the local favorite
states in energy are a quaternary and four types of pairs.
The quaternary is an instantaneous state which may separate into two pairs
randomly because there is no difference in energy between them.
The hoping terms in the Hamiltonian split the degeneracy to form a band of
charge-density wave states where the quaternary and pairs move
from site to site. As is known  in the absence of orbital
degree of freedom \cite{Emery} 
that the spin-density excitation turns over a spin
to break a pair at a cost $|U|$ in energy. The spin-orbital-density 
excitation, however, turns over either a spin or a orbital which
results in three basic processes. In addition to the process
of breaking a pair at the cost $|U|$ in energy, one process transmits
a favorite pair into an unfavorite pair at the cost $2|U|$ in energy.
The process of breaking a quaternary into a trinity and
a single costs $|U|$. Breaking a quaternary into
two unfavorite pairs will cost $4|U|$, which is not a basic
process because it can be represented by two processes of the second type.
All the other complicated processes can always be represented as a 
composition of those three basic processes. Thus we believe that there are
three elementary quasi-particles involved in the excitations
in the spin-orbital sector.

For the case (ii), the favorite states in energy are two types of trinitys
and three types of pairs. The charge-density wave states arising 
from the hopping terms are of the movement of the trinitys and pairs 
from site to site. The spin-orbital-density excitation that turns over
either a spin or a orbital involves in three basic processes which cost
$|U|$, $2|U|$ and zero in energy. Therefore, there are three elementary 
quasi-particles, in which a gap-less node is expected to exist.   

With the help of the above discussions we are now in the position to employ
elementary degenerate perturbation theory for more quantitative formulations.
It is sufficient for calculating the low temperature properties 
to consider the lowest band only. For those two types of partially attractive 
on-site couplings that were discussed previously, 
all electrons remain in either
the favorite pairs, trinitys, or quaternary in the lowest band. 
Because it breaks pairs, trinitys, or quaternary, the perturbation part
(the hopping term) has vanishing first order matrix elements. 
Thus it must be calculated to second order by considering the virtual
transitions into the next band. After some algebra one gets,
\begin{eqnarray}
(\varepsilon-\varepsilon_0)a_{g}&=&\sum_{g'}<g|H^2_t|g'>a_{g'}
  \nonumber\\
a_g&=&\frac{< g |H_t|\psi>}{\varepsilon-\varepsilon_0}
\end{eqnarray}
where $|g>$ denote various degenerate states of the 
unperturbed ground states, i.e., $H_U|g>=\varepsilon_0|g>$. 
$H_t$ stands for the hopping term and $H_U$ for the interaction term
of Eq. (\ref{eq:Hamiltonian}). 
After the similar calculation as in 
\cite{Emery}, we obtain the effective Hamiltonians that will
be helpful for studying repulsive model.

\section{Effective models of strong repulsive coupling}\label{sec:effect}

We now study the repulsive model $U_{ab}=U>0$. First we consider the 
half-filled band ($N=2L$). For the ground state in this case, 
every sites are doubly occupied by electrons. 
The excitation of charge-density 
waves brings about inevitably a triple occupancy of site at a cost
of $|U|$ in energy at least. 
By making a particle-hole canonical transformation:
\begin{eqnarray}
C_i(x)\mapsto C_i^+(x),\; i=2,4,
  \nonumber\\
C_j(x)\mapsto C_j(x),\;  j=1,3, 
  \nonumber
\end{eqnarray}
the Hamiltonian becomes
\begin{eqnarray}
&\tilde{H}&=t\displaystyle\sum_{\stackrel{a}{<x,x'>}}
           (-1)^a C_a^+(x)C_a(x')\nonumber\\
&+&U\sum_{\stackrel{a<a'}{x}}(-1)^{a+a'}n_a(x)n_{a'}(x)
            +\sum_{x,a}V_a (n_a(x)+\frac{1}{2}),
\label{eq:tildeHamiltonian}
\end{eqnarray}
where $V_a =[1-3(-1)^a]U/2$.
Clearly, the repulsive on-site coupling becomes the 
partially attractive case (i) which we already discussed. 
In the unperturbed states, the sites originally occupied by 
$|\tdown\bdown>$ become empty, whereas 
those occupied by $|\tup\bup>$ are replaced by 
$|\tup\tdown\bup\bdown>$. The other four kinds of double occupancies
exchanged their positions. The degenerate perturbation theory 
can now be used in the same way as it was done in the partially 
attractive models that we considered before. 
After reversing the canonical transformation, 
we obtain the effective Hamiltonian as follows
\begin{eqnarray}
H_{\rm eff}&=&\displaystyle\frac{t^2}{|U|}
       \sum_{<x,x'>}\left[h(x,x')-\frac{3}{4}\right],\nonumber\\
h(x,x')&=&\sum_{mn}g^{mn}O_m(x)O_n(x')
        +\sum_{\alpha\in\Delta}E_\alpha(x)E_{-\alpha}(x').
\label{eq:effectHamiltonian}
\end{eqnarray}
where $\Delta$ denotes the set of roots of the $A_3$ Lie algebra,
$E_\alpha(x)$ and $O_m(x)$ are generators of the Lie algebra
given by $E_{\alpha_m}(x)=C^+_m(x)C_{m+1}(x)$, 
$O_m(x)=C^+_m(x)C_m(x)-C^+_{m+1}(x)C_{m+1}(x)$,
$m=1,2,3$.
These operators generate the whole states from the highest weight state
$|\tup\tdown>$ in the following way,
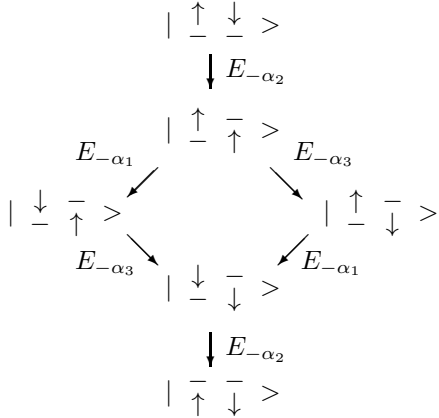
\begin{figure}
\setlength{\unitlength}{1mm}   
\begin{picture}(66,55)(-2,0)
\linethickness{0.5pt}
\put(26,52){$\mid\tup\tdown>$}
\put(32,49){\vector(0,-1){4.5}}\put(34,46){$E_{-\alpha_2}$}
\put(26,38){$\mid\tup\bup>$}
\put(25,34){\vector(-1,-1){4}}\put(14,35){$E_{-\alpha_1}$}
\put(40,34){\vector(1,-1){4.2}}\put(43,35){$E_{-\alpha_3}$}
\put(5,27){$\mid\tdown\bup>$}
\put(47,27){$\mid\tup\bdown>$}
\put(21,25){\vector(1,-1){4}}\put(14,20.5){$E_{-\alpha_3}$}
\put(45,25){\vector(-1,-1){4}}\put(44,20.5){$E_{-\alpha_1}$}
\put(26,17){$\mid\tdown\bdown>$}
\put(32,12){\vector(0,-1){4.5}}\put(34,9){$E_{-\alpha_2}$}
\put(26,3){$\mid\bup\bdown>$}
\end{picture}
\end{figure}
\noindent
where the $x$ in $E_\alpha(x)$ is omitted.
Those states form the vector representation of $A_3$
(isomorphic to $D_3$)
Lie algebra. Whence it is a SO(6) Heisenberg model. 

As to the case of quarter-filling (N=L) with repulsive
on-site coupling, the ground state has one electron on each site.
The excitation of charge-density waves brings about double 
occupancy of sites indispensably and costs energy of $|U|$ at least. 
It is again interesting to make a canonical transformation
\begin{eqnarray}
C_4(x)\mapsto C_4^+(x),\; C_l(x)\mapsto C_l(x), \;\; l=1,2,3.\nonumber
\end{eqnarray}
Consequently, the sites at first occupied by $|\bdown>$
are now empty whereas the other sites are doubly occupied,
i.e., by either ${\small |\tup\bdown>}$, ${\small |\tdown\bdown>}$,
or $|\bup\bdown>$. The repulsive on-site coupling 
becomes the partially attractive case (ii) which we discussed previously. 
Again, the degenerate perturbation theory can be applied as before. 
The effective Hamiltonian is obtained as a SU(4) Heisenberg
model. The local states form the spinor representation
of $A_3$ Lie algebra, and their relations in the weight diagram were 
given \cite{LiMSZ} by:
\begin{figure}
\setlength{\unitlength}{1mm}  
\begin{picture}(66,12)(0,0)
\linethickness{0.5pt}
\put(0,2){$\mid\tup>$}
\put(12,2){\vector(1,0){6}}\put(12,4){$E_{-\alpha_1}$}
\put(21,2){$\mid\tdown>$}
\put(33,2){\vector(1,0){6}}\put(33,4){$E_{-\alpha_2}$}
\put(42,2){$\mid\bup>$}
\put(54,2){\vector(1,0){6}}\put(54,4){$E_{-\alpha_3}$}
\put(63,2){$\mid\bdown>$}
\end{picture}
\end{figure}
\noindent
The evaluation of correlation function in the 
strong coupling limit will be considerably simple by means of the 
corresponding effective Hamiltonians. 

\section{Symmetry breaking}\label{sec:breaking}
 
The underlying SU$_d$(4) symmetry is fulfilled for the isotropic
on-side coupling $U_{a a'}=U$. There is no phase separation between 
spin wave and orbital wave at the SU(4) point. A complete 
separation is expected to occur after the anisotropic 
on-side couplings in the spin-orbital configuration are introduced. 
Since the `diagonal' coupling $U_{a a}$ has no physical contribution 
due to fermionic
wave function vanishes when two electrons at the same site being 
the same SU(4) state (i.e., a=a'), we can introduce six parameters 
$v_{a b}$ $(a<b)$ to break down the SU(4) symmetry, i.e.,
$U_{a a'}=U+v_{a a'}$.

It is not difficulty to find the possible symmetry breakings
by calculating the commutation relations between the Hamiltonian and the SU(4)
generators. There are two ways to break the SU(4) down to the
 SU(3)$\times$U(1). 
For $v_{12}=v_{13}=v_{23}$, $v_{14}=v_{24}=v_{34}$
(or $v_{12}=v_{13}=v_{14}$, $v_{23}=v_{24}=v_{34}$),
the residue symmetry SU(3)$\times$U(1) is generated by
$\{ O_1, O_2, E_{\pm\alpha_1}, E_{\pm(\alpha_1+\alpha_2)},
E_{\pm\alpha_2},O_3\}$   
(or by $\{ O_1, O_2, O_3, E_{\pm\alpha_2}, E_{\pm(\alpha_2+\alpha_3)} 
E_{\pm\alpha_3}\}$).
This is a two-parameter hierarchy. For a three-parameter
hierarchy, $v_{13}=v_{23}=v_{14}=v_{24}$, the residue symmetry is 
SU(2)$\times$SU(2)$\times$U(1) generated by 
$\{ O_1, E_{\pm\alpha_1}, O_3, E_{\pm\alpha_3}, O_2\}$.
When $v_{12}=v_{13}$ and $v_{24}=v_{34}$, the residue symmetry becomes
SU(2)$\times$U(1)$\times$U(1) with 
$\{O_2, E_{\pm\alpha_2}, O_1, O_3\}$
as its generators.
This is obviously a four-parameter hierarchy.
Furthermore, if either $v_{12}\neq v_{13}$ or $v_{24}\neq v_{34}$
the previous SU(2)$\times$U(1)$\times$U(1) will be
broken to the U(1)$\times$U(1)$\times$U(1) generated by
$\{O_1,O_2,O_3\}$.

As the SU(4) Lie algebra is of rank 3 (there are three generators 
in its Cartan subalgebra) the Zeemann-like interactions
with external fields reads
\[
H_Z=\sum_{x,m}h_m O_m(x),
\]
where $m=1,2,3$.
In general, it breaks the symmetry down to the minimum residue
symmetry  U(1)$\times$U(1)$\times$U(1).
However, if $h_2=h_3=0$, it regains 
a SU(2)$\times$U(1)$\times$U(1) generated by
$\{O_1, O_2, O_3, E_{\pm\alpha_3}\}$. 
Similarly another SU(2)$\times$U(1)$\times$U(1) generated by
$\{O_1, E_{\pm\alpha_1} O_2, O_3\}$
regains for $h_1=h_2=0$

Since it is the SU(4) singlet, the ground state is invariant under any 
SU(4) rotations. Except for the singlet-excitation and 
the pure charge-excitation states which are still of invariant under 
SU(4) rotation,  the other multiplet-excitation states vary.
For the system with $N=4n$, there is always an axis in flavor space
along which the multiplet-excitation states are invariant 
under a rotation of $2\pi$.   

\section{Discussions}\label{sec:discussions}

We have shown both the underlying and hidden symmetries of 
Hubbard model with orbital degeneracy. 
We derived the effective Hamiltonian in strong repulsive coupling
for both half-filled and quarter-filled band.
The band is half-filled at all temperature if the chemical potential
is $3U/2$. It is shown that there are three elementary modes
involved in the excitations in the spin-orbital sector.

We did not specify the dimensions in the above discussions on the 
symmetries of the system and its low-temperature effective Hamiltonians. 
It is interesting to consider one dimensional case \cite{LiGYE} 
because the exact solution
in one dimension always provides non-perturbative features. 
The exact solution can be obtained by means of Bethe-Yang ansatz
similar to ref.\cite{Sutherland68} if states with site occupation of more
than two are excluded \cite{ChoyHaldane}.
Although the multi-component generalizations of Hubbard model 
in one dimension were
explored by several authors\cite{Schlottmann,Choy} 
in various aspects, the relationship between
the enlarged internal degree of freedom and concrete physics was not
clearly exhibited. 
It is convenient to take thermodynamics limit to the Bethe ansatz
equation by introducing density distributions of the quasi-momentum
$k$ and that of three rapidities for the spin-orbital double (we
call SU(4) flavor). 
The ground state being of real roots and no holes
is a SU(4) singlet, accordingly, both spin and orbital
are in ``anti-ferromagnetic'' states.
The features of low-lying excitations are
studied by considering the contributions of holes and 
complex 2-strings. 
An important consequence is that the excitations in charge
sector and flavor sector are separated. In charge sector the elementary
modes are holon and anti-holon (particle), and the real excitations
are gap-less particle-holon and gapful holon-holon excitations etc.
There are three types (in agreement with the above general analyses in 
any dimension) of flavorons as the elementary excitation modes
in flavor sector,
namely, two quadruplets carrying spin 1/2 and orbital 1/2 that form
respectively the fundamental representation and the conjugate 
representation of SU(4); a hexaplet carrying either spin 1 or orbital 1
which forms a 6-dimensional representation moreover. These
flavorons compound to constitute real excitations. 
The details of one dimensional model are given in another paper\cite{LiGYE}.

We also analyzed the possible symmetry breakings caused either 
by extending the model to anisotropic on-side coupling or  
by introducing external field. Our analyses based on the  
Hamiltonian structure will be helpful for further discussions 
on the phase diagram by means of numerical DMRG. 

YQL acknowledges the supports of AvH-Stiftung, NSFC-19975040, and EYFC98, 
beneficial discussions with M. Ma and F.C. Zhang are also acknowledged. 

\end{multicols}

\begin{thebibliography}{99}

\bibitem{Nagaosa}
T. Tokura and N. Nagaosa, Science {\bf 288}, 462 (2000).

\bibitem{McWh}
D.B. McWhan, T.M. Rice, and J.P. Remeika,
Phys. Rev. Lett {\bf 23}, 1384 (1969);
D.B. McWhan, A. Menth, J.P. Remeika, W.F. Brinkman, and T.M. Rice,
Phys. Rev. B {\bf 7}, 1920 (1973); 
C. Castellani, C.R. Natoli, and J. Ranninger,
Phys. Rev. B {\bf 18}, 4945, 4967 and 5001 (1978).

\bibitem{Word}
R.E. Word, S.A Werner, W.B. Yelon, J.M. Honig, and S. Shivashankar,
Phys. Rev. B {\bf 23}, 3533 (1981);
K. I. Kugel and D. I. Khomskii, 
Usp. Fiz Nauk. {\bf 136}, 621 (1982). 

\bibitem{Bao}
W. Bao, C. Broholm, S.A. Carter, T.F. Rosenbaum, G. Aeppli,
S.F. Trevino, P. Metcalf, J.M. Honig, and J. Spalek, 
Phys. Rev. Lett. {\bf 71}, 766 (1993).

\bibitem{Pen}
H.F. Pen, J. Brink, D.I. Komomskii, and G.A. Sawatzky,
Phys. Rev. Lett. {\bf 78}, 1323 (1996);
B.J. Sternlieb, J. P. Hill, and U.C. Wildgruber,
Phys. Rev. Lett. {\bf 76}, 2196 (1996);
Y. Moritomo, A. Asamitsu, H. Kuwahara, and Y. Tokura,
Nature {\bf 380}, 141 (1996).
 
\bibitem{Feiner}
L.F. Feiner, A.M. Oles, and J. Zaanan, 
Phys. Rev. Lett. {\bf 78}, 2799 (1997);
S. Ishihara, M. Yamanaka, N. Nagaosa, 
Phys. Rev. B {\bf 56}, 686 (1997);
R. Shiina, T. Nishitani, and H. Shiba,
J. Phys. Soc. Japan {\bf 66}, 3159 (1997);
K. Yamaura, M. Takano, A. Hirano, and R. Kanno, 
J. Solid State Chemistry {\bf 127}, 109 (1997);
H. Kawano, R. Kajimoto, H. Yoshizawa, Y. Tomioka, 
H. Kuwahara, and Y. Tokura,  
Phys. Rev. Lett. {\bf 76}, 2196 (1997);
Y. Okimoto, T. Katsufuji, T. Ishikawa, T. Arima, and Y. Tokura,   
Phys. Rev. B {\bf 55}, 4206 (1997).
 
\bibitem{Arovas}
D.P. Arovas and A. Auerbach, Phys. Rev. B {\bf 52}, 10114 (1995).

\bibitem{Marston}
A.Onufric and J.B. Marston, Phys. Rev. B {\bf 59}, 12573 (1999).

\bibitem{LiMSZ}
Y.Q. Li, M. Ma, D.N. Shi, and F.C. Zhang,
Phys. Rev. Lett. {\bf 81}, 3527(1998); 
{\it ibid} Phys. Rev. B {\bf 60}, 12781 (1999).
 
\bibitem{Ueda}
Y. Yamashita, N, Shibata, and K. Ueda, Phys. Rev. B {\bf 58}, 9114(1998).

\bibitem{Azaria}
P. Azaria, E. Boulat, and P. Lecheminant,
Phys. Rev. B {\bf 61}, 12112 (2000).

\bibitem{IQAffleck}
C. Itoi, S. Qin, and I. Affleck, Phys. Rev. B {\bf 61}, 6747 (2000).  ;
Y. Yamashita, N, Shibata, and K. Ueda, 
J. Phys. Soc. Jpn {\bf 69}, 242 (2000).

\bibitem{phasesep}
N.A. Babushkina, L.M. Belova, D.I. Khomskii, K.I. Kugel, O.Yu. Gorbenko, 
and A.R. Kaul, Phys. Rev. B {\bf 59}, 6994 (1999);
M. Urhara, S. Mori, C.H. Chen, and S.W. Cheong,  
Nature {\bf 399}, 560 (1999).


\bibitem{Emery}
V.J. Emery, Phys. Rev. B {\bf 14}, 2989(1976).

\bibitem{Schlottmann}
P. Schlottmann, Phys. Rev. B {\bf 43}, 3101 (1991).

\bibitem{Choy}
T.C. Choy, Phys. Lett. {\bf 80 A}, 49 (1980).

\bibitem{spin3/2}
Although the system of spin-3/2 fermions is a four-component system,
for which one can also introduce fermion creation operators
$C^+_n(x)$, ($n=1,2,...,4$), the whole local states can be generated by 
a single operator $S^-$ from the highest weight state $|3/2>$. 
The underlying symmetry is a $SU(2)$ with the following generators:
\begin{eqnarray}
S^z&=&\sum_x\sum_{n=1}^4(n-5/2)C^+_n(x)C_n(x),\nonumber\\
S^+&=&\sum_x\sum_{m=-1}^1\sqrt{4-|m|}C^+_{3+m}(x)C_{2+m}(x),\nonumber\\
S^-&=&(S^+)^\dagger,\nonumber
\end{eqnarray}
they obey $[S^+, S^-]=2S^z$.

\bibitem{Gilmore} {\it e.g.}
R. Gilmore, 
{\it Lie groups, Lie algebras and some of their applications},
John Wiley \& Sons, New York, 1974.

\bibitem{Assaraf}
R. Assaraf, P. Azaria, M. Caffarel, and P. Lecheminant, 
Phys. Rev. B 60, 2299 (1999)

\bibitem{chargeSU2}
M. Pernici, Europhys. Lett. {\bf 12}, 751 (1990);
H.J. Schulz, Phys. Rev. Lett. {\bf 65}, 2462 (1990);
S. Zhang, Phys. Rev. Lett. {\bf 65}, 120 (1990).

\bibitem{Aligia}
A.A. Aligia and L. Arrachea, Phys. Rev. B 60, 15332 (1999).
A.A. Zvyagin and P. Schlottmann, Phys. Rev. B {\bf 60}, 6292 (1999).

\bibitem{LiGYE}
Y.Q. Li, S.J. Gu, Z.J. Ying, and U. Eckern, 
Phys. Rev. B {\bf 62}, 4866 (2000).

\bibitem{Sutherland68}
B. Sutherland, Phys. Rev. Lett. {\bf 20}, 98 (1968); 
{\it ibid}, Phys. Rev. B {\bf 12}, 3795 (1975);
E.H. Lieb and F.Y. Wu, Phys. Rev. Lett. {\bf 20}, 1445 (1968).

\bibitem{ChoyHaldane}
T.C. Choy and F.D.M. Haldane, Phys. Lett. A {\bf 90}, 83 (1982).

\end{thebibliography}
\end{document}